%% file: image_matching_daniel.tex
\documentclass[11pt]{article}
\usepackage{amsmath}
\usepackage{epsfig}
\usepackage{graphicx}
\usepackage{textcomp}
\usepackage{fullpage}

\input{pream}

\begin{document}

\title{Tight Approximation of Image Matching}

\author{
    Simon Korman \\
    School of EE \\
    Tel-Aviv University  \\
    Ramat Aviv, {\sc Israel} \\
    {\tt simon.korman@gmail.com}
 \and Daniel Reichman \\
    Faculty of Math and CS \\
    Weizmann Institute of Science  \\
    Rehovot, {\sc Israel} \\
    {\tt daniel.reichman@gmail.com}
 \and Gilad Tsur \\
     Faculty of Math and CS \\
    Weizmann Institute of Science  \\
    Rehovot, {\sc Israel} \\
    {\tt gilad.tsur@gmail.com}
}

\newcommand{\calN}{{\cal{N}}}
\newcommand{\calZ}{{\cal{Z}}}
\newcommand{\oneNsquared}{\{1,\dots,n\}^2}
\newcommand{\oneNcubed}{\{1,\dots,n\}^3}
\newcommand{\aff}{\bar{A}}
\maketitle
\begin{abstract}

In this work we consider the {\em image matching} problem for two grayscale $n \times n$ images, $M_1$ and $M_2$ (where pixel values range from $0$ to $1$). Our goal is to find an affine transformation $T$ that maps pixels from $M_1$ to pixels in $M_2$ so that the differences over pixels $p$ between $M_1(p)$ and $M_2(T(p))$ is minimized. Our focus here is on sublinear algorithms that give an approximate result for this problem, that is, we wish to perform this task while querying as few pixels from both images as possible, and give a transformation that comes close to minimizing the difference.

We give an algorithm for the image matching problem that returns a transformation $T$ which minimizes the sum of differences (normalized by $n^2$)  up to an additive error of $\epsilon$  and performs $\tilde{O}(n/\epsilon^2)$ queries. We give a corresponding lower bound of $\Omega(n)$ queries showing that this is the best possible result in the general case (with respect to $n$ and up to low order terms).

In addition, we give a  significantly better algorithm for a natural family of images, namely, smooth images. We consider an image smooth when the total difference between neighboring pixels is $O(n)$. For such images we provide an approximation of the distance between the images to within an additive error of $\epsilon$ using a number of queries depending polynomially on $1/\epsilon$ and not on $n$. To do this we first consider the image matching problem for $2$ and $3$-dimensional {\em binary} images, and then reduce the grayscale image matching problem to the $3$-dimensional binary case.

\end{abstract}

\newpage

\section {Introduction}

Similarity plays a central part in perception and categorization of visual stimuli. It is no wonder that similarity has been intensely studied, among others, by cognitive psychologists \cite{Tver, HCR} and computer vision and pattern recognition researchers.
Much of the work on computer vision, including that on image matching, involves algorithms that require a significant amount of processing time, whereas many of the uses of these algorithms would typically require real-time performance.

A motivating example is that of image registration~\cite{ZF, Szel}. Here we are given two images of a particular scene or object (e.g., two pictures taken from a video sequence) and wish to match one image to the other, for tasks such as motion detection, extraction of $3$-dimensional information, noise-reduction and super-resolution. Many advances were made in dealing with this task and it can now be performed in a wide variety of situations. However, image registration algorithms are generally time consuming.

Image registration is an application of a more abstract computational problem - the image matching problem~\cite{HL, HL2, LV}. In this problem we are given two digital $n \times n$ images $M_1$ and $M_2$ and wish to find a transformation that changes $M_1$ so that it best resembles $M_2$. In this work we consider the distance between two $n \times n$ images $M_1$ and $M_2$ when we perform {\em affine} transformations on their pixels. Namely, given an affine transformation $T$, we sum over all pixels $p$ the absolute value of the  difference between $M_1(p)$ and $M_2(T(p))$, where the difference is considered to be $1$ for pixels mapped outside $M_2$.
The distance between  $M_1$ and $M_2$ is defined as the minimum such distance taken over all affine transformations.
Our focus is on affine transformations as such transformations are often used when considering similarity between images. We limit ourselves to trasformations with a bounded scaling factor. This is congruent with applications, and prevents situations such as one image mapping to very few pixels in another. Exact algorithms for this problem generally enumerate all possible different transformations, fully checking how well each transformation fits the images.
Hundt and Li{\'s}kiewicz~\cite{HL} give such an algorithm for the set of affine transformations  on images with $n \times n$ pixels (transformations on which we focus in this paper), that runs in time $\Theta(n^{18})$. They also prove a lower bound of $\Omega(n^{12})$ on the number of such transformations (which implies a $\Omega(n^{12})$ lower bound on algorithms using this technique).

As known exact algorithms have prohibitive running times, image registration algorithms used in practice are typically heuristic. These algorithms often reduce the complexity of the problem by roughly matching ``feature points"~\cite{ZF, Szel} - points in the images that have relatively distinct characteristics. Such heuristic algorithms are not analyzed rigorously, but rather evaluated empirically.

Two related problems are those of shape matching and of point set matching, or point pattern matching. In shape matching~\cite{Velt} the goal is to find a mapping $T$ between two planar shapes $S_1$ and $S_2$, minimizing a variety of distance measures between the shapes $T(S_1)$ and $S_2$. A problem of similar flavor is that of point set matching ~\cite{GIMV}, where we are given two (finite) sets of points $A$ and $B$ in a Euclidean space and seek to map $A$ to a set $T(A)$ that minimizes the distance between $T(A)$ and $B$ under some distance metric. Algorithms for these exact problems were give by Alt et al'~\cite{alt1988congruence} and by Chew et al'~\cite{Kleinberg_Huttenlocher} both require prohibitive running times. Recent research ~\cite{GIMV} has focused on finding transformations that are close to the optimal one requiring less time. It should be noted that the running times the algorithms in ~\cite{GIMV} are superlinear in the number of points in $A$ and $B$. We emphasize that these works are concerned with planar shapes and point sets rather than digital images.

Our main contribution is devising \emph{sublinear} algorithms for the image matching problem. Sublinear algorithms are extremely fast (and typically randomized) algorithms that use techniques such as random sampling to asses properties of objects with arbitrarily small error. The number of queries made by such algorithms is sublinear in the input size, and generally depends on the error parameter. The use of sublinear algorithms in image processing was advocated by Rashkodnikova~\cite{Ras} who pioneered their study for visual properties.  She gave algorithms for binary ($0-1$) images, testing the properties of connectivity, convexity and being a half-plane. In her work, an image is considered far from having such a property if it has a large hamming distance from every image with the property. Ron and Tsur~\cite{RT:focs} introduced a different model that allowed testing of sparse images (where there are $o(n^2)$ different $1$-pixels) for similar properties. Kleiner et al. \cite{KKNB} give results on testing images for a partitioning that roughly respects a certain template. Unlike the aforementioned works we do not deal only with binary images, but also consider grayscale images, where every pixel gets a value in the range $[0,1]$.

\subsection{Our Results}

In this work we prove both general results and results for {\em smooth} images.

\BE

\item {\em General Upper Bound:} We present an algorithm that when given access to any two $n \times n$ grayscale images $M_1$ and $M_2$ and a precision parameter $\epsilon$
returns a transformation $T$ such that the distance between $M_1$ and $M_2$ using $T$ is at most $\epsilon$ greater than the minimum distance between them (taken over all affine transformations). The query complexity of this algorithm is $\Theta(n/\epsilon^2)$, which is sublinear in $n^2$, the size of the matrices.

\item {\em Lower Bound:} We show that every algorithm estimating matching between images within additive error smaller than $1/4$ must make an expected $\Omega(n)$ number of queries.

\item {\em Upper Bound For Smooth Images:} We show that if the images $M_1$ and $M_2$ are smooth, that is, for both images the total difference between neighboring pixels is $O(n)$,  then for every positive $\epsilon$ we can find a transformation $T$ such that the distance between $M_1$ and $M_2$ using $T$ is at most $\epsilon$ greater than the minimum distance between them. This can be done using a number of queries that is polynomial in $1/\epsilon$ and does not depend on $n$.

\EE

Being smooth is a property of many natural images - research has shown a power-law distribution of spatial frequencies in images~\cite{SH, MAH}, translating to very few fast changes in pixel intensity. While we show that our algorithm works well with smooth images, we note that distinguishing between images that have a total difference between neighboring pixels of $O(n)$ and those with a total of $O(n)+k$ requires $\Omega(n^2/k)$ queries.

An unusual property of the way distance between images is defined in this work is that it is not symmetric. In fact, an image $M_1$ may have a mapping that maps all its pixels to only half the pixels in $M_2$, so that each pixels is mapped to a pixels with the same value, while any mapping from $M_2$ to $M_1$ leaves a constant fraction of the pixels in $M_2$ mapped either outside $M_1$ or to pixels that do not have the same color (To see this consider an image $M_1$ that has only black points, and an image $M_2$ that is black on one the left side and white on the other). We note that one can use the algorithms presented here also to measure symmetric types of distances by considering inverse mappings.

\paragraph{Techniques}

\subparagraph{The Algorithm for the General Case: }
Imagine that sampling a pair of pixels, $p \in M_1$ and $q \in M_2$, would let us know how well {\em each} affine transformation $T$ did with respect to the pixel $p$, that is, what the difference is between $M_1(p)$ and $M_2(T(p))$. The way we define grayscale values (as ranging from $0$ to $1$), we could sample $\Theta(\epsilon^2)$ random pairs of points and have, for every transformation, an approximation of the average difference between points up to an additive error of $\epsilon$ with constant probability. As there are polynomially many different affine transformations if we increased the number of samples to amplify the probability of correctness, we could use $\tilde{O}(\log(n)/\epsilon^2)$ queries and return a transformation that was $\epsilon$-close to the best\footnote{The $\tilde{O}$ symbol hides logarithmic factors}. However, when we sample $p \in M_1$ and $q \in M_2$ uniformly at random we get a random pixel and its image under only a few of the different transformations. We show that $\tilde{O}(n/\epsilon^2)$ queries suffice to get a good estimation of the error for all interesting transformations (that is, transformations that map a sufficiently large portion of pixels from $M_1$ to pixels in $M_2$). Using these pixels we can return a transformation that is close to optimal as required.

\subparagraph{The Lower Bound: } We prove the lower bound by giving two distributions of pairs of images. In the first, the images are random $0-1$ images and far from each other. In the second, one image is partially created from a translation of the other. We show that any algorithm distinguishing between these families must perform $\Omega(n)$ expected queries.
The proof of the lower bound is somewhat similar to the lower bound given by Batu et al.~\cite{BEKMRRS} on the number of queries required  to approximate edit distance. Here we have a two-dimensional version of roughly the same argument. Note that a random $0-1$ image is far from being smooth, that is, many pixels have a value significantly different from that of their neighbors.

\subparagraph{The Algorithm For Smooth Images:}
Our analysis of the algorithm for smooth images begins by considering binary images. The {\em boundary} of a $0-1$ image $M$ is the set of pixels that have a neighboring pixel with a different value.
We consider two affine transformations $T, T'$ close if for every pixel $p$ the distance in the plane between $T(p)$ and $T'(p)$ is small. Only points that are close to the boundary might be mapped to different values by close transformations $T$  and $T'$ (meaning that the pixel will be mapped correctly by one and not by the other - see Figure~1). It follows that if there is a big difference in the distance between $M_1$ and $M_2$ when mapped by $T$ and their distance when mapped by $T'$, then the {\em perimeter}, the size of the boundary, is large. This implies that when the perimeter is small, one can sample a transformation $T$ and know a lot about the distance between images for transformations that are ``close" to $T$. This idea can be generalized to $3$-dimensional binary images. Such $0-1$ images are a natural object in $3$ dimensions as color is not a feature typically attributed to areas within a body. More importantly, however, one can use $3$-dimensional binary images to model $2$-dimensional grayscale images. Smooth grayscale images, i.e., images where the sum of difference (in absolute value) between neighboring pixels is $O(n)$, translate to $3$-dimensional binary images that have a small perimeter. An appropriate version of the $3$-dimensional algorithm can be used to get a good approximation for the mapping between two grayscale images.

\paragraph{Organization}

We begin by giving some preliminaries in Section~\ref{app.sec.prelim}.
We then describe and prove the correctness of the algorithm for the general case (with a query complexity of $\tilde{O}(n/\epsilon^2)$). We give the lower bound in the next section. Following that we give the algorithm for smooth binary images in Section~\ref{app.sec.alg}. In Section~\ref{app.sec.construction} we give an explicit construction of an $\epsilon$-net of transformations
such that any transformation is close to one of the those in the net. In Section~\ref{3D.sec} we give the three-dimensional version of our algrithm, and in Section~\ref{grayscale.sec} we show how to use this version to work with grayscale images.


\section{Preliminaries}\label{app.sec.prelim}

We are given two images represented by $n\times n$ matrices. For grayscale images the values of entries in the matrix are in the range $[0,1]$ and for binary images they are either $0$ or $1$.

\BD
A {\em pixel} $p$ in an $n\times n$ image $M$ is a pair of coordinates, namely a pair $(i,j) \in \oneNsquared$. We denote this as $p \in M$.
\ED

\BD
The {\em value} of a pixel $p = (i,j)$ in an image $M$ is $M[i,j]$, or $M(p)$.
\ED

\BD
For $r\in {\cal{R}}^2$ we denote by $\floor{r}$ the pixel $p$ that the point $r$ falls in.
\ED

\BD
A transformation $T$ has a {\em scaling factor} in the range $[1/c,c]$ (for a positive constant $c$) if for all vectors $v$ it holds that $||v||/c \leq||Tv|| \leq c||v||$.
\ED

Here we are particularly interested in affine transformations in the plane that are used to map one pixel to another, when these transformations
have a scaling factor in the range $[1/c,c]$ for a fixed positive constant $c$. Such a transformation $T$ can be seen as multiplying the pixel vector by a $2\times 2$ non-singular matrix
and adding a "translation" vector, then rounding down the resulting numbers.
When comparing two images, requiring the matrix to be non-singular prevents the transformation from mapping the image plane in one image onto a line or a point in the other.

Given an affine transformation in the form of a matrix $A$ and a translation vector $t$, there is a corresponding transformation $T(p) = \floor{Ap + t}$. We call $T$ an {\em image-affine} transformation and we say that $T$ is {\em based} on $A$ and $t$. Generally speaking, when we discuss algorithms getting an image-affine transformation as input, or enumerating such transformations, we assume that these transformations are represented in matrix format. 


\BD
\label{def.distance}
The distance between two $n \times n$ images $(M_1, M_2)$ with respect to a transformation $T$, which we denote $\Delta_T(M_1,M_2)$, is defined as

$$\frac{1}{n^2} \Big[ |\{p \in M_1~|~T(p) \notin M_2 \}| + \sum_{p\in M_1 | T(p)\in M_2} |M_1(p) - M_2(T(p)) | \Big]$$

\ED

Note that the distance $\Delta_T(M_1,M_2)$ ranges from $0$ to $1$.

\BD
We define the {\em Distance} between two images $(M_1, M_2)$ (which we denote $\Delta(M_1,M_2)$) as the minimum over all image-affine transformations $T$ of $\Delta_T(M_1,M_2)$.
\ED

\BD
Two different pixels $p=(i,j)$ and $q=(i',j')$ are adjacent if $|i-i'|\leq 1$ and $|j-j'|\leq 1$.
\ED

The following definitions relate to binary ($0-1$) images:

\BD
A pixel $p = (x,y)$ is a {\em boundary} pixel in an image $M$ if there is an adjacent pixel $q$ such that $M(p) \neq M(q)$.
\ED

\BD
The {\em perimeter} of an image $M$ is the set of boundary pixels in $M$ as well as the $4n-4$ outermost pixels in the square image. We denote the size of the perimeter of $M$ by $P_M$.
\ED
\
Note that $P_M$ is always $\Omega(n)$ and $O(n^2)$.

\section{The General Case}

We now present the algorithm for general images. The lower bound we will prove in Section~\ref{app.sec.LB} demonstrates that this algorithm has optimal query complexity, despite having a prohibitive running time. The main signficance of the algorithm is in showing that one can achieve query complexity of $\tilde{O}(n)$.  It is an open question if one can achieve this query complexity in sublinear time or even significantly faster than our running tme.

\subsection{The Algorithm}

\begin{figure}[h]
\BA\label{alg.general.case}
Input: Oracle access to $n\times n$ images $M_1, M_2$, and a precision
parameter $\epsilon$.
\BE
\item Sample $k = \tilde{\Theta}(n/\eps^{2})$ pixels ${\cal{P}} =
p_1,\dots, p_k$ uniformly at random (with replacement) from $M_1$.
\item Sample $k$ pixels ${\cal{Q}} = q_1,\dots, q_k$ uniformly at
random (with replacement) from $M_2$.
\item Enumerate all image-affine transformations $T_1,\dots, T_m$
(Recall that $m$, the number of image-affine transformations, is in
$O(n^{18})$).
\item For each transformation $T_\ell$ denote by $Out_\ell$ the number
of pixel coordinates that are mapped by $T_\ell$ out of the region
$[1, n]^2$.
\item For each transformation $T_\ell$ denote by $Hit_\ell$ the number
of pairs $p_i, q_j$ such that $T_\ell(p_i) = q_j$, and denote by
$Bad_\ell$  the value  $\frac{1}{|\{p\in {\cal{P}} , q\in {\cal{Q}} | T_\ell(p_i) =
q_j\}|}\sum_{p_i, q_j \in \{p\in {\cal{P}} , q\in {\cal{Q}} | T_\ell(p_i) = q_j\}} |M_1(p_i)
- M_2(q_j)|$
\item Return $T_\ell$ that minimizes $(n^2-Out_\ell) \cdot Bad_\ell$
(discarding transformations $T_\ell$ such that $Hit_\ell < \epsilon$).
\EE
\EA
\end{figure}

\BT\label{thm.alg.general.case.good}
With probability at least $2/3$ Algorithm~\ref{alg.general.case}
returns a transformation $T$ such that $|\Delta_{T}(M_1, M_2) -
\Delta(M_1, M_2)| < \epsilon$.
\ET

We prove Theorem~\ref{thm.alg.general.case.good} by showing that for
any fixed transformation $T_\ell$ (where $Hit_\ell \geq \epsilon$) the
sample we take from both images gives us a value $Bad_\ell$ that is a
good approximation of the value $\frac{1}{|\{p \in M_1, q\in M_2|
T_\ell(p_i) = q_j\}|}\sum_{p_i, q_j \in \{p \in M_1, q\in M_2|
T_\ell(p_i) = q_j\}} |M_1(p_i) - M_2(q_j)|$  with high probability,
and applying a union bound. To show this we give several definitions
and claims. For these we fix an image-affine transformation $T$ and
two images $M_1$ and $M_2$, so that $T$ maps at least $\epsilon/2$ of
the points in $M_1$ to points in $M_2$ (note that transformations that
do not map such an $\epsilon/2$ portion of pixels are discarded by the
algorithm with very high probability).

\BE
\item Let $T(M_1)$ be the set of pixels $q \in M_2$ such that there
exist pixels $p \in M_1$ so that $T(p) = q$.
\item For a set of pixels $Q\in M_2$ let $T^{-1}(Q)$ denote the set
$\{p\in M_1 | T(p)\in Q\}$.
\item We denote by ${\cal{Q}}'$ the points that are in ${\cal{Q}}$
(the sample of points taken from $M_2$) and in $T(M_1)$.
\item We denote by ${\cal{P}}'$ the points $p \in {\cal{P}}$ such that
$T(p) \in {\cal{Q}}'$.
\item For a pixel $q \in M_2$ we denote by $|q|$ the number of pixels
$p\in M_1$ such that $T(p) = q$.
\item For a pixel $q \in M_2$ we denote by $\hat{q}$ the sum over
pixels $p\in M_1$ such that $T(p) = q$ of $|M_1(p) - M_2(T(p))|$.
\item Denote by $p_{bad}$ the average over pixels $p$ from those
mapped from $M_1$ to $T(M_1)$ of $|M_1(p) - M_2(T(p))|$.
\item Denote by $\hat{p}_{bad}$ the value $(\sum_{q\in{\cal{Q}}'}
\hat{q})/(\sum_{q\in{\cal{Q}}'} |q|)$.
\EE

\BCM\label{clm.general.alg.enough.points.in.sample}
With probability at least $1/(8n^{18})$ over the choice of ${\cal{P}}$
and ${\cal{Q}}$
the size of ${\cal{Q}}'$ is $\tilde{\Omega}(n/\eps)$ and the size of
${\cal{P}}'$ is $\tilde{\Omega}(\log(n)/\eps^3)$.
\ECM

\BPF
The probability of any particular pixel in ${\cal{Q}}$ belonging to
${\cal{Q}}'$ is at least $\epsilon/2$, and ${\cal{Q}}$ is of size
$\tilde{\theta}(n/\epsilon^2)$ (where pixels are chosen
independently). Hence the expected number of points in ${\cal{Q}}'$
is $\Omega(n/\eps)$.
An additional factor of $\Theta(\log(n))$ hidden in the $\tilde{\Theta}$ notation of $k$ assures us (using Chernoff
bounds) that the probability ${\cal{Q}}'$ not being large enough is at
most $1/(8n^{18})$ as required.

Assume the first part of the claim holds. Recall that no more than a
constant number of pixels from $M_1$ are mapped to any pixel in $M_2$,
and therefore $|T^{-1}({\cal{Q}}')| = \Omega(n/\eps)$. As the pixels
of ${\cal{P}}$ are chosen independently and uniformly at random from
the $n^2$ pixels of $M_1$, each pixel in ${\cal{P}}$ is mapped to a
pixel in ${\cal{Q}}'$ with a probability of $\Omega(1/(n\epsilon))$.
Hence, the expected size of ${\cal{P}}'$ is $\Omega(1/\eps^3)$ and the
second part of the claim follows (via a similar argument).
\EPF

\BCM
With probability at least $1/(8n^{18})$ over the choice of ${\cal{P}}$
and ${\cal{Q}}$ it holds that
$|\hat{p}_{bad} - p_{bad}| < \epsilon/4$.
\ECM

\BPF
Note that $p_{bad}$ equals $(\sum_{q \in T(M_1)} \hat{q}) / (\sum_{q
\in T(M_1)} |q|)$. Now, consider the value $\hat{p}_{bad} =
(\sum_{q\in{\cal{Q}}'} \hat{q})/(\sum_{q\in{\cal{Q}}'} |q|)$.
Each pixel $q \in {\cal{Q}}'$ is chosen uniformly at random and
independently from the pixels in $T(M_1)$. To see that the claim holds
we note that with probability at least $1 - 1/(8n^{18})$ (using
Hoeffding  bounds and the fact that with high probability ${\cal{P}}'$ is $\tilde{\Omega}(\log(n)/\eps^3)$) we have that
$| (\sum_{q\in{\cal{Q}}'} \hat{q} / |{\cal{Q}}'|) - (\sum_{q\in
T(M_1)} \hat{q} / |T(M_1)|)| = \O(\epsilon)$
 and that $| (\sum_{q\in{\cal{Q}}'} |q| / |{\cal{Q}}'|) - (\sum_{q\in
T(M_1)} |q| / |T(M_1)|)| = \O(\epsilon)$. The claim follows.
\EPF

\BCM
With probability at least $1/(8n^{18})$ over the choice of ${\cal{P}}$
and ${\cal{Q}}$ it holds that
$|Bad_\ell - \hat{p}_{bad}| < \epsilon/2$
\ECM

\BPF
We have that $\hat{p}_{bad} = \frac{\sum_{q\in{\cal{Q}}'}
\hat{q}}{\sum_{q\in{\cal{Q}}'} |q|}$. It follows that $\hat{p}_{bad}$
equals $E_{p \in T^{-1}({\cal{Q}}')} [M_1(p) = M_2(T(p))]$ where $p$
is chosen uniformly at random from $T^{-1}({\cal{Q}}')$. The pixels in
 ${\cal{P}}'$ are chosen uniformly at random from $T^{-1}({\cal{Q}}')$
and (with sufficiently high probability), by
Claim~\ref{clm.general.alg.enough.points.in.sample} there are
$\tilde{\Omega}(\log(n)/\eps^3)$ such pixels. The claim follows using
Hoeffding bounds.
\EPF

We thus see that $Bad_\ell$ is $\epsilon$-close to the average difference between $M_1(p)$ and $M_2(T_\ell(p))$ for pixels $p \in M_1$ mapped by $T_\ell$ to $M_2$. As $Out_\ell$ is exactly the number of pixels mapped by $T_\ell$ out of $M_2$, the claim follows from the definition of distance.

\subsection{Lower Bound}\label{app.sec.LB}

We build a lower bound using binay images, and parameterize the lower bound with the image perimeter (see Theorem~\ref{app.thm.lower.bound}). Note that for an image $M$ having a perimeter of size $k$ implies that the total difference between pixels and their neighbors in $M$ is $\Theta(k)$. We use a bound on this total difference to discuss the smoothness of grayscale images later in this work.

\BT\label{app.thm.lower.bound}
Fix $k > 0$ such that $k = o(n)$ ($k$ may depend on $n$). Let $A$ be an algorithm that is given access to pairs of images $M_1,M_2$ where $max(P_{M_1},P_{M_2}) = \Omega(n^2/k)$. In order to distinguish with high probability between the following two cases:
\BE
\item $\Delta(M_1,M_2) < 4/16$
\item $\Delta(M_1,M_2) > 7/16$
\EE
 $A$ must perform $\Omega(n/k)$ queries.
\ET

In order to prove Theorem~\ref{app.thm.lower.bound} we will first focus on the case where $k = 1$. Namely, we show that any algorithm that satisfies the conditions as stated in the theorem for all images with $max(P_{M_1},P_{M_2}) = \Theta(n^2)$ must perform $\Omega(n)$ queries. Following this we will explain how the proof extends to the case of a general $k$.

We use Yao's principle - we give two distributions ${\calD}_1, {\calD}_2$ over pairs of images  such that the following holds:
\BE
\item $\Pr_{(M_1,M_2)\texttildelow {\calD}_1}[\Delta(M_1,M_2) > 7/16] > 1-o(1)$
\item $\Pr_{(M_1,M_2)\texttildelow {\calD}_2}[\Delta(M_1,M_2) < 4/16] = 1$
\EE
  and show that any deterministic algorithm that distinguishes with high probability between pairs drawn from ${\calD}_1$ and those drawn from ${\calD}_2$ must perform $\Omega(n)$ expected queries. We now turn to describe the distributions.

The distribution ${\calD}_1$ is the distribution of pairs of images where every pixel in $M_1$ and every pixel in $M_2$ is assigned the value $1$ with probability $0.5$ and the value $0$ otherwise, independently. Pairs in the distribution ${\calD}_2$ are constructed as follows.
$M_1$ is chosen as in ${\calD}_1$. We now choose uniformly at random two values $s_h,s_v$ ranging each from $0$ to $n/8$. Pixels $(i,j)$ in  $M_2$ where $i < s_h$ or $j < s_v$ are chosen at random as in ${\calD}_1$. The remaining pixels $(i,j)$ satisfy $M_2(i,j) = M_1(i - s_h, j - s_v)$. Intuitively, the image $M_2$ is created by taking $M_1$ and shifting it both horizontally and vertically, and filling in the remaining space with random pixels.

Both distributions ${\calD}_1$ and ${\calD}_2$ possess the required limitation on the size of the boundaries (i.e. $max(P_{M_1},P_{M_2}) = \Theta(n^2)$). It suffices to show this with respect to the image $M_1$, which is constructed in the same way in both distributions. It is easy to see (since $M_1$'s pixels are independently taken to be 0 or 1 uniformly at random) that $\Pr[P_{M_1}\geq \frac{n^2}{4}] = 1 - o(1)$.

We now proceed to prove the properties of ${\calD}_1$ and ${\calD}_2$. Starting with ${\calD}_2$, given the transformation $T$ that shifts points by $s_v$ and $s_h$, all but a $15/64$'th fraction (which is smaller than $1/4$) of $M_1$'s area matches exactly with a corresponding area in $M_2$ and the following claim holds.

\BCM
$\Pr_{(M_1,M_2)\texttildelow {\calD}_2}[\Delta(M_1,M_2) < 4/16] = 1$
\ECM

In order to prove that pairs of images drawn from ${\calD}_1$ typically have a distance of at least $7/16$ we first state the following claim~\cite{HL}:
\BCM
The number of image-affine transformations $T$ between $n \times n$ images $M_1$ and $M_2$ that map at least one of $M_1$'s pixels into $M_2$ is polynomial in $n$.
\ECM

\BCM\label{app.clm.D1.pairs.far}
$\Pr_{(M_1,M_2)\texttildelow {\calD}_1}[\Delta(M_1,M_2) > 7/16] > 1-o(1)$
\ECM

\BPF
Consider two images $M_1, M_2$ that are sampled from ${\calD}_1$. The value $\Delta_T(M_1, M_2)$ for an arbitrary transformation $T$ is $\Delta_T(M_1, M_2) \leq \Pr_{p\in M_1}[T(p)\in M_2 \wedge M_1(p) = M_2(T(p))]$.
 For any pixel $p$, over the choice of $M_1$ and $M_2$, it holds that $\Pr [T(p)\in M_2 \wedge M_1(p) = M_2(T(p))] \leq 1/2$ (if $T(p)\in M_2$ the probability is $1/2$). The random (over the choice of $M_1, M_2$)  variable  $\Delta_T(M_1, M_2)$ has an expectation of at most $n^2/2$. As it is bounded by the sum of $n^2$ independent $0-1$ random variables with this expectation, the probability that  $\Delta_T(M_1, M_2) < (1/2-\epsilon)n^2$ for any positive fixed $\epsilon$ is $\Theta(e^{-n^2})$. As $\Delta(M_1, M_2) = \min_{T}\Delta_T(M_1, M_2)$, and as there are at most a polynomial number of transformations $T$, the claim follows using a union bound.
\EPF

The proof of Theorem~\ref{app.thm.lower.bound} for the case $k=1$ is a consequence of the following claim.

\BCM
Any algorithm that given a pair of $n \times n$ images $M_1, M_2$
acts as follows:
\BE
\item Returns $1$ with probability at least $2/3$ if $\Delta(M_1, M_2) \leq 4/16$.
\item Returns $0$ with probability at least $2/3$ if $\Delta(M_1, M_2) \geq 7/16$.
\EE
must perform $\Omega(n)$ expected queries.
\ECM

\BPF
To show this we consider any deterministic algorithm that can distinguish with probability greater than $1/2+\epsilon$  (for a constant $\epsilon >0$) between the distributions ${\calD}_1$ and ${\calD}_2$. Assume (toward a contradiction) such an algorithm $A$ that performs $m = o(n)$ queries exists. We will show that with very high probability over the choice of images, any new query $A$ performs is answered independently with probability $0.5$ by $0$ and with probability $0.5$ by $1$. This implies the Theorem.

The fact that any new query $A$ performs is answered in this way is obvious for ${\calD}_1$ - here the pixels are indeed chosen uniformly at random.

We now describe a process $P$ that answers a series of $m$ queries performed by $A$ in a way that produces the same distribution of answers to these queries as that produced by pairs of images drawn from ${\calD}_2$. This will complete the proof. The process $P$ works as follows (we assume without loss of generality that $A$ never queries the same pixel twice):

\BE
\item\label{app.set.answers} Select $m$ bits $r_1,\dots,r_m$ uniformly and independently at random. These will (typically) serve as the answers to $A$'s queries.
\item Select uniformly at random two values $s_h,s_v$ ranging each from $0$ to $n/8$.
\item For $q_k = (i , j)$ - the $k$'th pixel queried by $A$, return the following:
\BE
	\item\label{app.bad.1} If $q_k$ is queried in $M_1$, and $M_2(i + s_h, j + s_v)$ was sampled, return $M_2(i + s_h, j + s_v)$.
	\item\label{app.bad.2} If $q_k$ is queried in $M_2$, and $M_1(i - s_h, j - s_v)$ was sampled, return $M_1(i - s_h, j - s_v)$.
	\item Else, return $r_k$.

\EE
\EE

Obviously, $P$ has the same distribution of answers to queries as that of images drawn from ${\calD}_2$ - the choice of $s_h, s_v$ is exactly as that done when selecting images from ${\calD}_2$, and the values of pixels are chosen in a way that respects the constraints in this distribution.
 We now show that the probability of reaching Steps~\ref{app.bad.1} and $\ref{app.bad.2}$ is $o(1)$. If $P$ does not reach these steps it returns $r_1,\dots, r_m$ and $A$ sees answers that were selected uniformly at random. Hence, the claim follows.

Having fixed $r_1,\dots,r_m$ in Step~\ref{app.set.answers}, consider the queries $q'_1,\dots, q'_m$ that $A$ performs when answered $r_1,\dots, r_{m-1}$ (that is, the query $q'_1$ is the first query $A$ performs. If it is answered by $r_1$ it performs the query $q'_2$, etc.). In fact, we will ignore the image each query is performed in, and consider only the set of pixel locations $\{p_k = (i_k, j_k)\}$. Step~\ref{app.bad.1} or $\ref{app.bad.2}$ can only be reached if a pair of pixels $p_k, p_\ell$ satisfies $|i_k - i_\ell| = s_v$ and  $|j_k - j_\ell| = s_h$. There are $\Theta(m^2)$ such pairs, and as $m = o(n)$ we have that $m^2 = o(n^2)$. As the number of choices of $s_v, s_h$ is in $\Theta(n^2)$, the probability of $s_v, s_h$ being selected so that such an event occurs is $o(1)$ as required.
\EPF

\paragraph{Proving Theorem~\ref{app.thm.lower.bound} for the Case $k>1$ (sketch).}
We construct distributions similar to those above, except that instead of considering single pixels we partition each image to roughly $n^2/k^2$ blocks of size $k \times k$, organized in a grid. The distributions ${\calD}_1, {\calD}_2$ are similar, except that now we assign the same value to all the pixels in each block. For the distribution ${\calD}_2$ we select $s_v,s_h$ to shift the image by multiples of $k$. The remainder of the proof is similar to the case where $k=1$, but the number of queries that an algorithm must perform decreases from $\Omega(n)$ to $\Omega(n/k)$, while the boundary size decreases from $\Theta(n^2)$ to  $\Theta(n^2/k)$.

\section{The Smooth Image Case}\label{bounded.prerim.sec}

\subsection{The Algorithm for Binary Images with Bounded Perimeter}\label{app.sec.alg}

Given a pair of binary images $M_1, M_2$ with $P_{M_1} = O(n)$ and $P_{M_2} = O(n)$ our approach to finding an image-affine transformation $T$ such that $\Delta_T(M_1, M_2) \leq \Delta(M_1, M_2) + \epsilon$ is as follows. We iterate through a set of image affine transformations that contain a transformation that is close to optimal (for all images with a perimeter bounded as above), approximating the quality of every transformation in this set. We return the transformation that yields the best result.

We first show (in Claim~\ref{app.clm.compare.correct}) how to approximate the value $\Delta_T(M_1,M_2)$ given a transformation $T$. We then show that for two affine transformations $\bar{T}, \bar{T}'$ that are close in the sense that for every point $p$ in the range $\oneNsquared$ the values $\bar{T}(p)$ and $\bar{T}'(p)$ are not too far (in Euclidean distance), the following holds. For the {\em image affine} transformations $T, T'$ based on $\bar{T}$ and $\bar{T}'$ the values $\Delta_T(M_1, M_2)$ and $\Delta_{T'}(M_1, M_2)$ are close. This is formalized in Theorem~\ref{app.thm.small.distance.small.change} and Corollary~\ref{app.cor.exists.close.pair}. Finally we claim that given a set $\cal{T}$ of affine transformations such that for every affine transformation there exists a transformation close to it in $\cal{T}$, it suffices for our purposes to check all image affine transformations based on transformations in $\cal{T}$. In Section~\ref{app.sec.construction} we give the construction of such a set.

The following claims and proofs are given in terms of approximating the distance (a numerical quantity) between the images. However, the algorithm is constructive in the sense that it finds a transformation that has the same additive approximation bounds as those of the distance approximation.

\BCM\label{app.clm.compare.correct}
Given images $M_1$ and $M_2$ of size $n\times n$ and an image-affine transformation $T$, let $d = \Delta_T(M_1,M_2)$. Algorithm~\ref{app.alg.approx.dist.by.trans} returns a value $d'$ such that $|d'-d| \leq \epsilon$ with probability $2/3$ and performs $\Theta(1/\epsilon^2)$ queries.
\ECM

\begin{figure}[h]
\BA\label{app.alg.approx.dist.by.trans}
Input: Oracle access to $n\times n$ images $M_1, M_2$, precision parameter $\epsilon$ and a transformation $T$ (given as a matrix and translation vector).
\BE
\item Sample $\Theta(1/\epsilon^2)$ values $p \in M_1$. Check for each $p$ whether $T(p)\in M_2$, and if so check whether $M_1(p) = M_2(T(p))$.
\item Return the proportion of values that match the criteria $T(p)\in M_2$  and $M_1(p) = M_2(T(p))$.
\EE
\EA
\end{figure}

The approximation is correct to within $\epsilon$ using an additive Chernoff bound.

We now define a notion of distance between affine transformations (which relates to points in the plane):

\BD
Let $\bar{T}$ and $\bar{T}'$ be affine transformations. The $l^n_\infty$ distance between $\bar{T}$ and $\bar{T}'$ is defined as $\underset{p\in [1,n+1)^2}{\operatorname{max}} \|\bar{T}(p)-\bar{T}'(p)\|_2$.
\ED

The notion of $l^n_\infty$ distance simply quantifies how far the mapping of a point in an image according to $\bar{T}$ may be from its mapping by $\bar{T}'$. Note that this definition doesn't depend on the pixel values of the images, but only on the mappings $\bar{T}$ and $\bar{T}'$, and on the image dimension $n$.

The following fact will be needed for the proof of Theorem~\ref{app.thm.small.distance.small.change}.

\BCM \label{app.perimeter_bound_lemma}
Given a square subsection $M$ of a binary image and an integer $b$, let $\hat{P}_M$ denote the number of boundary pixels in $M$. If $M$ contains at least $b$ $0$-pixels and at least $b$ $1$-pixels, then $\hat{P}_M \geq \sqrt{b}$.
\ECM

\BPF
Let $M$ be a square of $d \times d$ pixels. Note that $d > \sqrt{b}$. To see the claim holds we consider three cases. In the first case, all rows and all columns of $M$ contain both $0$ and $1$ pixels. In such a case each row contains at least one boundary pixel, $\hat{P}_M \geq d > \sqrt{b}$, and we are done. In the second case there exists, without loss of generality, a row that does not contain the value $0$, and all columns contain the value $0$. Again this means there are at least $d$ boundary pixels (one for each column), $\hat{P}_M \geq d > \sqrt{b}$, and we are done. Finally, consider the case that there are both rows and columns that do not contain the value $0$. This means that there is a boundary pixel for each row and for each column that {\em do} contain the value $0$. If there were fewer than $\sqrt{b}$ boundary pixels this would mean there are fewer than $\sqrt{b}$ rows and columns that contain $0$ pixels, and $M$ could not contain $b$ different $0$ pixels. This would lead to a contradiction, and thus $\hat{P}_M \geq \sqrt{b}$, and we are done.
\EPF

\vspace{4pt}
We now turn to a theorem that leads directly to our main upper-bound results.

\BT\label{app.thm.small.distance.small.change}
Let $M_1,M_2$ be  $n\times n$ images and let $\delta$ be a constant in $(0,\sqrt 2)$. Let $T$ and $T'$ be image affine transformations based on the affine transformations $\bar{T}, \bar{T}'$, such that $l^n_\infty(\bar{T},\bar{T}')<\delta n$. It holds that $$\Delta_{T'}(M_1,M_2) \leq \Delta_T(M_1,M_2) + O\Big(\frac{\delta P_{M_2}}{n}\Big)$$
\ET
{

\BPF

The distance
$\Delta_T(M_1,M_2) = \frac{1}{n^2} \Big|\{p \in M_1~|~T(p) \notin M_2 \vee M_1(p) \neq M_2(T(p)) \}\Big|$ is composed of two parts. The first is the portion of pixels from $M_1$ that $T$ maps out of $M_2$. The second is the portion of pixels in $M_1$ that $T$ maps to pixels that have a different value in $M_2$. We will bound $\Delta_{T'}(M_1,M_2) - \Delta_{T}(M_1,M_2)$ . This amounts to bounding the change in the two values mentioned above.

We begin by bounding the number of pixels from $M_1$ that are mapped by $T$ to pixels of $M_2$ but aren't mapped to such pixels by $T'$.  As $l^n_\infty(\bar{T},\bar{T}')<\delta n$, all such pixels are at most $\delta n$-far from the outermost pixels of the image. We will bound the number of such pixels by $O(\delta n^2)$. Since $P_{M_2} > n$ (for it contains all the outermost pixels in $M_2$) and since we normalize by $n^2$, these pixels contribute $O(\frac{\delta P_{M_2}}{n})$ as required. We restrict the remaining analysis to pixels that have at least a distance of $\delta n$ from the outermost pixels in the image.

The second value can be viewed as the number of \emph{new} mismatches between $M_1$ and $M_2$ that are introduced, when replacing $T$ by $T'$ (which is not very different from $T$), and we will discuss this change (see Figure~1). Formally, if we denote this amount by $mis_{T\rightarrow T'} = |\{p\in M_1 | M_1(p) = M_2(T(p))\neq M_2(T'(p))\}|$, it would suffice to show that $$mis_{T\rightarrow T'} = O(\delta n P_{M_2})$$ (the amount of mismatches is normalized by $n^2$, the size of the image, in order to get the difference).
We will bound the amount of new mismatches by breaking the image $M_2$ into a grid of $\delta n \times \delta n$ squares ($1/\delta$ such squares on each dimension), showing how the contribution of each square to $mis_{T\rightarrow T'}$ depends on its contribution to the perimeter of the image $M_2$. For integers $i$ and $j$, both between $1$ and $1/\delta$, let $b_{i,j}$ be the $\delta n \times \delta n$ square located at the $i$th row and $j$th column of the squares grid defined above. Summing on these squares, we can write: $$mis_{T \rightarrow T'} = \sum_{i=1}^{1/\delta}\sum_{j=1}^{1/\delta}|\{p\in M_1 | T(p)\in b_{i,j}, M_1(p) = M_2(T(p))\neq M_2(T'(p))\}|$$

We now give several additional definitions:
\begin{itemize}
  \item Let $mis_{T\rightarrow T'}^{i,j} = |\{p\in M_1 | T(p)\in b_{i,j}, M_1(p) = M_2(T(p))\neq M_2(T'(p))\}|$ be the contribution of $b_{i,j}$ to $mis_{T\rightarrow T'}$. This definition implies that:
       \begin{itemize}
         \item $mis_{T\rightarrow T'} = \sum_{i=1}^{1/\delta}\sum_{j=1}^{1/\delta}mis_{T\rightarrow T'}^{i,j}$
         \item For any $i$ and $j$, $mis_{T\rightarrow T'}^{i,j}$ is an integer in the range $[0,\delta^2 n^2]$
       \end{itemize}
  \item Let $B_{i,j}$ denote the $3\delta n \times 3\delta n$ square (a block of $3 \times 3$ original grid squares), containing the square $b_{i,j}$ in its center.
  \item Let $P_{M_2}^{i,j}$ be the number of pixels in the perimeter of $M_2$ that exists within the square $B_{i,j}$
\end{itemize}

It obviously holds that:
\begin{equation}\label{app.maximum_mismatch}
    \sqrt{mis_{T\rightarrow T'}^{i,j}}<\delta\cdot n
\end{equation}
\begin{equation}\label{app.perimeter_sums}
    P_{M_2} \geq \frac{1}{9} \cdot \sum_{i=2}^{\frac{1}{\delta} - 1}\sum_{j=2}^{\frac{1}{\delta} - 1} P_{M_2}^{i,j}
\end{equation}

Since $l^n_\infty(\bar{T},\bar{T}')<\delta n$, each pixel $p\in M_1$ that is mapped by $T$ into $b_{i,j}$ is certainly mapped by $T'$ into $B_{i,j}$. It follows that  $$mis_{T\rightarrow T'}^{i,j} = |\{p\in M_1 | T(p)\in b_{i,j}, T'(p)\in B_{i,j},  M_1(p) = M_2(T(p))\neq M_2(T'(p))\}|$$

$mis_{T\rightarrow T'}^{i,j}$ is the sum of pixels $p\in M_1$, which are either 0-pixels or 1-pixels. Assume, with out loss of generality, that there are more such 0-pixels. These pixels account for at least half the amount:

$$mis_{T\rightarrow T'}^{i,j} \leq 2\cdot |\{p\in M_1 | T(p)\in b_{i,j}, T'(p)\in B_{i,j},  0 = M_1(p) = M_2(T(p))\neq M_2(T'(p))\}|$$

This implies that there are at least $\frac{0.5}{f(c)}\cdot mis_{T\rightarrow T'}^{i,j}$ 0-pixels in $b_{i,j}$ and at least $\frac{0.5}{f(c)}\cdot mis_{T\rightarrow T'}^{i,j}$ 1-pixels in $B_{i,j}$ where $f(c)$ is a constant depending only on $c$ (since our scaling factors are within the range $[\frac{1}{c},c]$, $f(c)=O(c^2)$ pixels from $M_1$ are mapped to the same pixel in $M_2$). In particular, the larger square $B_{i,j}$ contains at least $\frac{0.5}{f(c)}\cdot mis_{T\rightarrow T'}^{i,j}$ 0-pixels as well as at least $\frac{0.5}{f(c)}\cdot mis_{T\rightarrow T'}^{i,j}$ 1-pixels.

Using Claim~\ref{app.perimeter_bound_lemma}, we can conclude that:
\begin{equation}\label{app.local_perimeter}
	P_{M_2}^{i,j} \geq \sqrt {\frac{0.5}{f(c)}\cdot mis_{T\rightarrow T'}^{i,j}}
\end{equation}
and using the bounds of equations (\ref{app.maximum_mismatch}) and then (\ref{app.perimeter_sums}) and (\ref{app.local_perimeter}), we can conclude that:

$$mis_{T\rightarrow T'} = \sum_{i=2}^{\frac{1}{\delta} - 1}\sum_{j=2}^{\frac{1}{\delta} - 1} mis_{T\rightarrow T'}^{i,j} \leq \delta \cdot n \cdot \sum_{i=1}^{1/\delta}\sum_{j=1}^{1/\delta} \sqrt {mis_{T\rightarrow T'}^{i,j}}            \leq 9\sqrt{2 f(c)}\delta n P_{M_2}$$
\EPF
}

\begin{figure}[h]
\begin{center}
\includegraphics[width=140mm]{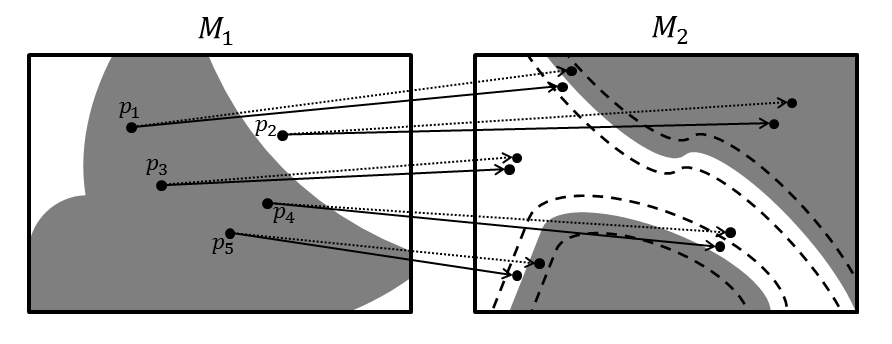}
\caption{\small
Consider two $\delta n-$close transformations between binary images $M_1$ and $M_2$ (the white/gray areas in the images correspond to 0/1 pixels). The solid and dotted arrows describe the action of the two transformations on the pixels $p_1,\dots,p_5$.
The areas between the dashed lines in $M_2$ contain the pixels that are $\delta n-$close to boundary pixels.
The only pixels in $M_1$ that are possibly mapped correctly by one transformation but not by the other are those that are mapped into the 'dashed' area by one of the transformations. In this example, only $p_1$ and $p_5$ are mapped correctly by one, but not by the other.
}

\end{center}
\label{app.fig.boundary}
\end{figure}







\BD
Let $\aff$ be the set of Image-Affine transformations. For a positive $\alpha$, the set of transformations ${\cal{T}}=\{T_i\}_{i=1}^l$ is an $\alpha$-cover of $\aff$ if for every $A$ in $\aff$, there exists some $T_j$ in ${\cal{T}}$, such that $l^n_\infty(A, T_j) \leq \alpha$.
\ED

We are going to show that for any given $n$ and $\delta >0$ there's a $\delta n$-cover of $\aff$ with size that does not depend on $n$ but
only on $\delta$.
Using this fact, given two
images $M_1$ and $M_2$ , we will run Algorithm~\ref{app.alg.approx.dist.by.trans} on every member of the cover, and get an approximation of $\Delta(M_1,M_2)$. In fact, we find a transformation $T \in \aff$ that realizes this bound.

\BCM \label{app.clm.approximating.transformation}
Let ${\cal{T}}=\{T_i\}_{i=1}^\ell$ be a $\delta n$-cover of $\aff$ and let $M_1, M_2$ be two $n\times n$ images. A transformation $T\in{\cal{T}}$ such that $$|\Delta_T(M_1,M_2) - \Delta(M_1,M_2)| \leq O(\frac{\delta}{n} \cdot max(P_{M_1}, P_{M_2}) + \epsilon)$$ can be found with high probability using $\tilde{O}(\ell/\epsilon^2)$ queries.
\ECM

\BPF
To find such a transformation $T$ we will run Algorithm~\ref{app.alg.approx.dist.by.trans} for $m = \Theta(\log{\ell})$ times on each of the $\ell$ transformations $\{T_i\}$ with precision parameter $\epsilon$, and set $s_i$ as the median of the results given for $T_i$.
 By the correctness of Claim~\ref{app.clm.compare.correct} and standard amplification techniques, for each $i$ the value $s_i$ will differ from $\Delta_{T_i}(M_1,M_2)$ by at most $\epsilon$ with probability at least $\frac{1}{3\ell}$ (we will say such a value $s_i$ is {\em correct}). Using a union bound we get that the probability of any $s_i$ not being correct is at most $1/3$. This will bound our probability of error, and from here on we assume all the values $s_i$ are indeed correct and show that we will get a transformation as required.

We now consider an image affine transformation $A$ such that $\Delta(M_1,M_2) = \Delta_A(M_1,M_2)$. By the fact that ${\cal{T}}$ is a $\delta n$-cover, there exists a  transformation $T_i$ such that $l^n_\infty(A, T_i) \leq \delta n$. Given Theorem~\ref{app.thm.small.distance.small.change} the value $\Delta_{T_i}(M_1,M_2) \leq \Delta(M_1, M_2) + O(\frac{\delta}{n} \cdot max(P_{M_1}, P_{M_2}))$ and thus the minimum value $s_j$ will not exceed $\Delta(M_1, M_2) + O(\frac{\delta}{n} \cdot max(P_{M_1}, P_{M_2}) + \epsilon)$. Choosing the transformation $T_j$ that this value is associated with, we get that $$|\Delta_{T_j}(M_1,M_2) - \Delta(M_1,M_2)| \leq O(\frac{\delta}{n} \cdot max(P_{M_1}, P_{M_2}) + \epsilon)$$ as required.
\EPF

\vspace{4pt}
In section \ref{app.sec.construction} we show the existence of a $\delta n$-cover of $\aff$ whose size is $\Theta(1/\delta^6)$. We can therefore conclude with the following corollaries:

\BCo
Given images $M_1,M_2$ and constants $\delta,\epsilon$, we have that $\Delta(M_1,M_2)$ can be approximated, using $\tilde{O}(1/\epsilon^2\delta^6)$ queries, with an additive error of $O(\frac{\delta}{n} \cdot max(P_{M_1}, P_{M_2}) + \epsilon)$.
\ECo

\BCo
Given images $M_1,M_2$ and constants $\delta,\epsilon$  such that $P_{M_1} = O(n)$ and $P_{M_2} = O(n)$, $\Delta(M_1,M_2)$ can be approximated, using  $\tilde{O}(1/\epsilon^2\delta^6)$ queries, with an additive error of $O(\delta + \epsilon)$.
\ECo



\subsection{Construction of a $\delta n$-cover of $\aff$} \label{app.sec.construction}

In this section we construct a $\delta n$-cover of $\aff$, which will be a product of several $1$-dimensional and $2$-dimensional grids of transformations, each covering one of the constituting components of a standard decomposition of Affine transformations~\cite{Hart}, which is given in the following claim.

\BCM \label{app.affine_trans_decomposition}
Every (orientation-preserving) affine transformation matrix $A$ can be decomposed into $A=T R_2 S R_1$, where $T, R_i, S$ are translation, rotation and non-uniform scaling matrices \footnote{arguments are similar for orientation-reversing transformations (which include reflection)}.
\ECM


We now describe a $6$-dimensional grid, which we will soon prove to be a $\delta n$-cover of $\aff$, as needed. According to claim \ref{app.affine_trans_decomposition}, every affine transformation can be composed of a rotation, scale, rotation and translation. These primitive transformations correspondingly have 1, 2, 1 and 2 degrees of freedom. These are: rotation angle, x and y scales, rotation angle and x and y translations. It is elementary, for instance, that if we impose a $2$-dimensional  grid of x and y translations, spaced in each direction by an interval of $\sqrt{2}\delta n$, then for any two neighboring translations $T_1$ and $T_2$ on this grid it holds that $l^n_\infty(T_1,T_2)<\delta n$. Since the range of possible translations is limited to the interval $[-n,n]$, the size of the $2$-dimensional  grid is $\Theta(1/\delta^2)$. Similarly for scaling, we are limited to the interval $[\frac{1}{c},c]$ and in order to have $l^n_\infty(S_1,S_2)<\delta n$ for neighboring scalings we use spacings of $\Theta(\delta)$. Likewise, we cover the $1$-dimensional  space of rotations, with angles in the interval $[0,2\pi]$ with spacings of $\Theta(\delta)$. Finally, by taking the cartesian product of these grids, we end up with a single grid of size $\Theta(1/\delta^6)$.

It remains to be shown that the grid we defined above, which we denote by $\cal{G}$, imposes a $\delta n$-cover of $\aff$.

\BCM \label{app.affine_distance_lemma}
For every $n$, for every $\delta'$, there exists a $\delta' n$-cover of $\aff$ of size $\Theta(1/\delta'^6)$.
\ECM

%

\BPF
Given the grid $\cal{G}$ and any image-affine transformation $A$, if we denote by $A'$ the nearest transformation to $A$ on the grid $\cal{G}$, we need to show that $l^n_\infty(A,A')<\delta n$. According to claim \ref{app.affine_trans_decomposition}, $A$ and $A'$ can be written in the form $A=T R_2 S R_1$ and $A'=T' R_2' S' R_1'$, such that $l^n_\infty(T,T')$, $l^n_\infty(R_1,R_1')$, $l^n_\infty(S,S')$ and $l^n_\infty(R_2,R_2')$ are all at most $\delta n$.

We now measure how differently $A$ and $A'$ might act on a pixel $p$, in order to obtain a bound on $l^n_\infty(A,A')$. At each stage we use the triangle inequality, accumulating additional distance introduced by each transformation as well as the $l^n_\infty$ bounds on the constituting transformations.

\BEQN
  \|S R_1(p) - S' R_1'(p)\| 
  &\leq& \|S (R_1p) - S' (R_1p)\| + \|S' (R_1p) - S' (R_1'p)\| \nonumber\\
  &=& \|(S-S') (R_1p)\| + \|S' (R_1p-R_1'p)\| \nonumber\\
  &\leq& \delta n + c\|R_1p-R_1'p\| = \delta n + c\delta n = (c+1)\delta n\nonumber
\EEQN

\BEQN
  \|R_2 S R_1(p) - R_2' S' R_1'(p)\| 
  &\leq& \|R_2 (S R_1p) - R_2' (S R_1p)\| + \|R_2 (S R_1p) - R_2' (S' R_1'p)\| \nonumber\\
  &=& \|(R_2-R_2') (S R_1p)\| + \|R_2' (SR_1p-S'R_1'p)\| \nonumber\\
  &\leq& \delta n + \|SR_1p-S'R_1'p\| = (c+2)\delta n\nonumber
\EEQN

\BEQN
  \|A(p)-A'(p)\| &=& \|T R_2 S R_1(p) - T' R_2' S' R_1'(p)\|\nonumber\\
  &\leq& \|T (R_2 S R_1p) - T'( R_2 S R_1p)\| + \|T' (R_2 S R_1p) - T' (R_2' S' R_1'p)\| \nonumber\\
  &=& \|(T-T') (R_2 S R_1p)\| + \|T' (R_2 S R_1p-R_2' S' R_1'p)\| \nonumber\\
  &\leq& \delta n + \|R_2 S R_1(p) - R_2' S' R_1'(p)\| = (c+3)\delta n\nonumber
\EEQN
The construction follows by setting $\delta = \delta'/(c+3)$.
\EPF



\subsection{3-Dimensional Images}\label{3D.sec}

In this section we generalize our techniques and results to $3$-dimensional images. 
One important application of the $3$-dimensional ($0-1$) setting is to the problem of aligning 3 dimensional solid objects, which are represented by $3$-dimensional $0-1$ matrices, where the objects are represented by the $1$s. The other motivation is the need to handle $2$-dimensional grayscale images. This is done in section~\ref{grayscale.sec}, where our algorithm is based on a reduction from grayscale images to $3$-dimensional $0-1$ images.

In this setting, we are given two images represented by $n\times n\times n$ ~$0-1$ matrices. The image entries are indexed by $voxels$, which are triplets in $\oneNcubed$ and the affine transformations in the 3-$d$ space act on a voxel by first multiplying it with a non-singular $3\times 3$ matrix $A$ (which accounts for rotation and anisotropic scale), then adding a 'translation' vector and finally rounding down to get a new voxel. The distance under a fixed affine transformation $T$ between two images $M_1, M_2$ is defined in an analogous way to the 2-dimensional case (definition \ref{def.distance}) and is denoted by $d = \Delta_T(M_1,M_2)$. So is the distance between two images with respect to affine transformations (that is $\Delta(M_1,M_2)$).  Voxels are considered $adjacent$ if they differ in each of their coordinates by at most 1 and a voxel is a $boundary$ voxel if it is adjacent to different valued voxels. Finally, the $perimeter$ of the image is the set of its boundary voxels together with its outer  $6n^2-12n+8$ voxels (and it is always $\Omega(n^2)$ and $O(n^3)$).

\vspace{12pt}
Given two images $M_1,M_2$ and an affine transformation $T$ we can approximate $\Delta_T(M_1,M_2)$ using the same methods as those used in Algorithm~\ref{app.alg.approx.dist.by.trans}. The only difference is that we sample voxels rather than pixels. Thus we have:
\BCM\label{app.clm.compare.correct.3d}
Given $3$-dimensional binary images $M_1$ and $M_2$ of size $n\times n \times n$ and an image-affine transformation $T$, let $d = \Delta_T(M_1,M_2)$. There is an algorithm that returns a value $d'$ such that $|d'-d| \leq \epsilon$ with probability $2/3$ and performs $\Theta(1/\epsilon^2)$ queries.
\ECM

Claim \ref{app.perimeter_bound_lemma} generalizes to the following:

\BCM \label{app.perimeter_bound_lemma_3D}
Given a cubic subsection $M$ of a binary 3-dimensional image with dimensions $h\times h \times h$, and an integer $b$, let $\hat{P}_M$ denote the number of boundary voxels in $M$. If $M$ contains at least $b$ $0$-voxels and at least $b$ $1$-voxels, then $\hat{P}_M = O(b^{2/3})$.
\ECM

\BPF
Assume without loss of generality that there are fewer $0$-voxels than $1$-voxels. We index the voxels in $M$ as $M(i,j,k)$. Let us denote by $M(i,j,\cdot)$ the sum $\sum_{k=1}^h M(i,j,k)$, and use $M(i,\cdot,k)$ and $M(\cdot,j,k)$ in a similar manner.
We first note several facts:

\BE
\item $h \geq (2b)^{1/3}$

\item The number of pairs $(i,j)$ such that $M(i,j,\cdot) > 0$ is at least $b^{2/3}$. This holds because there are at least $h^3/2$ different $1$-voxels. As each pair $(i,j)$ can account for at most $h$ different $1$-voxels, there must be at least $h^2/2 \geq b^{2/3}$ such pairs.

\item Either the number of pairs $(i,j)$ such that $M(i,j,\cdot) < h$ is at least $b^{2/3}$, or the number of pairs $(i,k)$ such that $M(i,\cdot,k) < h$
is at least $b^{2/3}$, or the number of pairs $(j,k)$ such that $M(\cdot,j,k) < h$ is at least $b^{2/3}$. This follows from the following claim which is a direct consequence of Lemma 15.7.5 in Alon and Spencer's book\cite{AS_probabilistic}:
\BCM
Consider a set $S$ of $b$ vectors in $S_1 \times S_2 \times S_3$. Let $S_{i,j}$ be the projection of $S$ into $S_i \times S_j$ (where $i \neq j$). If $b_{ij}=|S_{i,j}|$ then $b^2\leq \prod_{ij}b_{ij}$
\ECM
\EE

Assume without loss of generality that the number of pairs $(i,j)$ such that $M(i,j,\cdot) < h$ is at least $b^{2/3}$ and recall that the number of pairs $(i,j)$ such that $M(i,j,\cdot) >0$ is at least $b^{2/3}$. We consider two cases:

\BE
\item In the first case there are at least $b^{2/3}/2$ pairs of indices $(i,j)$ such that $0 < M(i,j,\cdot) < h$. Each such pair surely accounts for at least one boundary pixel, and we are done.
\item In the second case there are at least $b^{2/3}/2$ pairs of indices $(i,j)$ such that $M(i,j,\cdot) = 0$ and at least $b^{2/3}/2$ pairs of indices $(i,j)$ such that $M(i,j,\cdot) = h$. In this case one of the following will hold:
\BE
\item\label{caseI} There are at least $b^{1/3}/2$ indices $i$ such that there exists an index $j$ such that $M(i,j,\cdot) = 0$ and there are at least $b^{1/3}/2$ indices $i$ such that there exists an index $j$ such that $M(i,j,\cdot) = h$.
\item There are at least $b^{1/3}/2$ indices $j$ such that there exists an index $i$ such that $M(i,j,\cdot) = 0$ and there are at least $b^{1/3}/2$ indices $j$ such that there exists an index $i$ such that $M(i,j,\cdot) = h$.
\EE
  We assume without loss of generality that Case~\ref{caseI} holds. This means
that, again, one of two cases holds:
\BE
\item  There are more than $b^{1/3}/2$ indices $i$ such that there are both indices $j_0$ and $j_1$ such that $M(i,j_0,\cdot) = 0$ and $M(i,j_1,\cdot) = h$. In this case each such index accounts for $h$ boundary pixels, and we thus have at least $hb^{1/3}/2 \geq b^{2/3}/2$ boundary pixels, and we are done.

\item Otherwise, there is at least one index $i_0$ such that for all $j$ $M(i_0,j,\cdot) = 0$, and there is least one index $i_1$ such that for all $j$ $M(i_1,j,\cdot) = h$. But this means that for any pair of indices $(j,k)$ it holds that $M(i_0,j,k) = 0$ and $M(i_1,j,k) = 1$ and there must be at least one boundary pixel for each such pair $(j,k)$, giving us at least $h^2 \geq b^{2/3}$ boundary pixels and we are done.
\EE

\EE
\EPF

\vspace{12pt}

Our "central" theorem ~\ref{app.thm.small.distance.small.change} generalizes to the following:

\BT\label{app.thm.small.distance.small.change_3D}
Let $M_1,M_2$ be  $n\times n\times n$ images and let $\delta$ be a constant in $(0,\sqrt 3)$. Let $T$ and $T'$ be image affine transformations based on the affine transformations $\bar{T}, \bar{T}'$, such that $l^n_\infty(\bar{T},\bar{T}')<\delta n$. It holds that $$d_{T'}(M_1,M_2) \leq d_T(M_1,M_2) + O\Big(\frac{\delta P_{M_2}}{n}\Big)$$
\ET

\BPF (Outline of differences from the original proof)

The square grids $b$ and $B$ are now cubes of edge size $\delta n$ and $3\delta n$ respectively and are parametrized by the triplet $i,j,k$.

Some of our observations slightly change:
\begin{equation}\label{app.maximum_mismatch_3D}
    \sqrt[3]{mis_{T\rightarrow T'}^{i,j,k}}<\delta\cdot n
\end{equation}
\begin{equation}\label{app.perimeter_sums_3D}
    P_{M_2} \geq \frac{1}{27} \cdot \sum_{i=2}^{\frac{1}{\delta} - 1}\sum_{j=2}^{\frac{1}{\delta} - 1}\sum_{k=2}^{\frac{1}{\delta} - 1} P_{M_2}^{i,j,k}
\end{equation}

Using Claim~\ref{app.perimeter_bound_lemma_3D}, we can conclude that:
\begin{equation}\label{app.local_perimeter_3D}
	P_{M_2}^{i,j,k} \geq {(\frac{0.5}{f(c)}\cdot mis_{T\rightarrow T'}^{i,j,k}})^{2/3}
\end{equation}
and using the bounds of equations (\ref{app.maximum_mismatch_3D}) and then (\ref{app.perimeter_sums_3D}) and (\ref{app.local_perimeter_3D}), we can conclude that:

$$mis_{T\rightarrow T'} = \sum_{i=2}^{\frac{1}{\delta} - 1}\sum_{j=2}^{\frac{1}{\delta} - 1}\sum_{k=2}^{\frac{1}{\delta} - 1} mis_{T\rightarrow T'}^{i,j,k} \leq \delta \cdot n \cdot \sum_{i=1}^{1/\delta}\sum_{j=1}^{1/\delta}\sum_{k=1}^{1/\delta} ({mis_{T\rightarrow T'}^{i,j,k}})^{2/3}    \leq 27\sqrt{2 f(c)}\delta n P_{M_2}$$
\EPF

It is straightforward to extend the $2$-dimensional case and construct a $\delta n$ cover for the set of 3-dimensional affine transformations where the size of the cover depends only on $\delta$. As in the 2-dimensional case, the matrix $3\times 3$ matrix $A$ can be decomposed (using SVD decomposition) into a product of rotation, scaling and rotation matrices. Together with the final translation vector, we get a $\delta n-$cover of size $1/\delta^{10}$. The existence of such a cover along with a $3$-dimensional analog of claim \ref{app.clm.approximating.transformation} implies:

\label{corr.3.dimensional}
\BCo
Given 3-dimensional images $M_1,M_2$ and fixed constants $\delta,\epsilon > 0$  such that $P_{M_1} = O(n^2)$ and $P_{M_2} = O(n^2)$, the distance $\Delta(M_1,M_2)$ can be approximated, using  $\tilde{O}(1/\epsilon^2\delta^{10})$ queries, with an additive error of $O(\delta + \epsilon)$.
\ECo


\subsection{Grayscale Images}\label{grayscale.sec}

In this section we handle $2$-dimensional grayscale images by no longer limiting ourselves to binary $\{0,1\}$ valued pixels but rather allowing  a pixel $p$ to have any value $M(p)$ in the interval $[0,1]$. This model covers the commonly practiced discretizations (e.g. to 256 grey levels) of the intensity information in a digital image.

In the following definitions we extend the concept of the perimeter to grayscale images.

\BD
The \emph{gradient} of a pixel in a grayscale image $M$ is the maximal absolute difference between the pixel value and the pixel values of its adjacent pixels.
\ED

\BD
The \emph{perimeter size} $P_M$ of a grayscale image $M$ is defined as the sum of its pixels' gradients (where the gradient of each of the $4n-4$ outermost pixels of the image is counted as 1).
\ED

Notice, that the gradient of a pixel is a real valued number in $[0,1]$ and that if we consider a binary 0-1 image, its boundary pixels are exactly those with gradient one. Also, the perimeter size is $\Omega(n)$ and $O(n^2)$.

When dealing with binary $0-1$ images, our similarity measure between images was defined to be the maximal similarity between the images with respect to any Affine transformation on the image pixels. In the grayscale extension we would like to introduce further transformations, allowing our distance metric to capture (or be invariant to) illumination changes. That is, we would like to consider images that differ by a global linear change in pixel values to be similar. Such a linear change first multiplies all image pixels values by a 'contrast' factor $con$ and then adds to them a 'brightness' factor $bri$. As is custom in the field, pixel values that deviate from the $[0,1]$ interval as a result of such a transformation will be truncated so that they stay within the interval. Also, we limit $con$ to the interval $[1/c,c]$ for some positive constant $c$ and therefore $bright$ can be limited to $[-c,1]$ (since $con$ maps a pixel value into the range $[0,c]$). We denote the family of such intensity transformations by $\bar{\cal{BC}}$.



\BD
Let $T_1$ and $T_2$ be any two functions from $\bar{\cal{BC}}$. The $l^n_\infty$ distance between $T_1$ and $T_2$ is defined as the maximum over pixel values $v\in [0,1]$ of ${\operatorname{max}} \|T_1(v)-T_2(v)\|_2$ (which equals $\max |T_1(v)-T_2(v)|$).
\ED

We can now define the distance between grayscale images under a combination of an affine and an intensity transformation.

\BD
Let $T\in \aff$ be an Affine transformation and let $L\in\bar{\cal{BC}}$ be and intensity transformation. The distance between grayscale images $M_1, M_2$, with respect to $T$ and $L$ is:


$$\Delta_{T,L}(M_1,M_2) = \frac{1}{n^2} \sum_{p \in M_1} \big(1_{T(p) \notin M_2}+1_{T(p) \in M_2}\cdot| M_1(p) - L(M_2(T(p)))| \big)$$
\ED

We can now state our main result:

\BCM \label{app.clm.approximating.transformation.grayscale}

Given $n\times n$ grayscale images $M_1,M_2$ and positive constants $\delta$ and $\epsilon$, we can find transformations $T\in{\cal{T}}$ and $L\in{\cal{BC}}$ such that with high probability $$|\Delta_{T,L}(M_1,M_2) - \Delta(M_1,M_2)| \leq O\Big(\frac{\delta}{n} \cdot max(P_{M_1}, P_{M_2}) + \epsilon\Big)$$ using $\tilde{O}(1/\epsilon^2\delta^8)$ queries.
\ECM

\BPF
We will show Claim~\ref{app.clm.approximating.transformation.grayscale} holds by reducing the problem of approximating the distance between two $2$-dimensional grayscale images to that of approximating the distance between two $3$-dimensional $0-1$-images. In particular, we will map an $n\times n$ grayscale image $M$ to an $n\times n \times n$ binary image $M'$ defined as follows: $M'(i,j,k) = 1$ if and only if $M(i,j) \geq k/n$.

This essentially means that a pixel with intensity $g$ is represented by a column of pixels where the bottom $\floor{gn}$ pixels are $1$-pixels and the remaining are $0$-pixels.
The perimeter $P_{M'}$ of $M'$ is $\Theta(n^2)+P_{M} \cdot n$. This follows since a gradient of $g$ at a pixel $p$ creates $gn$ boundary pixels in $M'$. Any image-affine transformation $T$ of the grayscale image can be applied to a voxel in $M'$ without changing the voxels's third coordinate, that is, we can see the transformation $T$ as mapping between columns of voxels. Likewise, intensity transformations $L$ can be seen as applying only to the third coordinate of a voxel, that is, mapping pixels to higher or lower locations in their corresponding columns and truncating them to $n$ (1) if their value is larger (smaller) than $n$ (1). This truncation is equivalent to the truncation of the pixel values to the interval $[0,1]$ when applying intensity transformations on grayscale images.

We wish to derive a similar result to corollary \ref{corr.3.dimensional}. To do this, we consider a slightly different metric on 3-dimensional binary images. Namely, we limit the general family of 3-dimensional affine transformations to include only transformations that apply a two dimensional affine transformation on the first two coordinates as well as scale and translation on the third coordinate (which relate to the intensity component of the transformation). Call this family of transformations $\overline{S}$. Now we can proceed in a similar fashion to corollary \ref{corr.3.dimensional}.  Denote by $M_1'$ and $M_2'$ the resulting 3-dimensional images after applying our reduction on the 2-dimensional grayscale images $M_1$ and $M_2$. It holds that
$\Delta(M_1,M_2)=\Delta(M_1',M_2')$ as there is a one to one correspondence between transformations of grayscale images defined by a pair $T$ and $L$ between $M_1$ and $M_2$ and transformations in $\bar{S}$ between $M_1'$ and $M_2'$. Furthermore, by the way our reduction was defined, such a corresponding pair of transformations yield the same distance between both pairs of images.

We can now proceed along the same reasoning leading to corollary \ref{corr.3.dimensional}. Namely, we construct a $\delta n$ cover for our limited set of $3$-dimensional transformations. For the component of the 2-dimensional affine transformation we use the same cover used in section \ref{app.sec.construction} of size $\Theta(\frac{1}{\delta^6})$. For the intensity component we use a similar construction by dividing the scale and translation ranges in the third coordinate to step sizes of $\Theta(\delta n)$ and $\Theta(\delta)$ respectively. The resulting cover is of size $\Theta(\frac{1}{\delta^8})$ and it is easily shown to be a valid $\delta n$ cover. The assertion now follows in a similar way to corrolary \ref{corr.3.dimensional}. The only difference is that we consider only the set of restricted transformations $\overline{S}$ rather than the set of all 3-dimensional affine transformations.

%
\EPF

We conclude with the following corollary:

\BCo
Given $n \times n$ grayscale images $M_1,M_2$ and constants $\delta,\epsilon$  such that $P_{M_1} = O(n)$ and $P_{M_2} = O(n)$, the distance $\Delta(M_1,M_2)$ can be approximated, using  $\tilde{O}(1/\epsilon^2\delta^8)$ queries, with an additive error of $O(\delta + \epsilon)$.
\ECo

{\small
\bibliographystyle{plain}
\bibliography{test}
}

\end{document}

%% file: pream.tex
\renewcommand{\paragraph}[1]{{\protect\vspace{8pt}\noindent\sc{#1}}}

\newlength{\saveparindent}
\newlength{\saveparskip}
\newcommand{\BE}{\begin{enumerate}} \newcommand{\EE}{\end{enumerate}}
\newcommand{\BI}{\begin{itemize}} \newcommand{\EI}{\end{itemize}}
\newcommand{\BDes}{\begin{description}}\newcommand{\EDes}{\end{description}}
\newtheorem{alg}{Algorithm}      
\newcommand{\BA}{\begin{alg}} \newcommand{\EA}{\end{alg}}
\newtheorem{thm}{Theorem}[section]
\newcommand{\BT}{\begin{thm}} \newcommand{\ET}{\end{thm}}

\newtheorem{lem}{Lemma}[section]      
\newcommand{\BL}{\begin{lem}} \newcommand{\EL}{\end{lem}}

\newtheorem{clm}[lem]{Claim}            
\newcommand{\BCM}{\begin{clm}} \newcommand{\ECM}{\end{clm}}

\newtheorem{techcor}[thm]{Corollary} 
\newcommand{\BCo}{\begin{techcor}} \newcommand{\ECo}{\end{techcor}}

\newtheorem{cor}[thm]{Corollary}      
\newcommand{\BC}{\begin{cor}} \newcommand{\EC}{\end{cor}}
\newtheorem{prop}[thm]{Proposition}     
\newcommand{\BP}{\begin{prop}} \newcommand {\EP}{\end{prop}}
\newtheorem{conj} {Conjecture}      
\newcommand{\BCJ}{\begin{conj}} \newcommand{\ECJ}{\end{conj}}

\newtheorem{defn}{Definition}[section]         
\newcommand{\BD}{\begin{defn}} \newcommand{\ED}{\end{defn}}

\def\FullBox{\hbox{\vrule width 8pt height 8pt depth 0pt}}
\newcommand{\qed}{\;\;\;\FullBox}
\newenvironment{proof}{\noindent{\bf Proof:~~}}{\(\qed\)}
\newcommand{\BPF}{\begin{proof}} \newcommand {\EPF}{\end{proof}}
\newenvironment{proofof}[1]{\noindent{\bf Proof of {#1}:~~}}{\(\qed\)}
\newcommand{\BPFOF}{\begin{proofof}} \newcommand {\EPFOF}{\end{proofof}}
\newcommand{\qedsketch}{\;\;\;\Box}

\newenvironment{smallproof}{\noindent{\bf Proof:~~}}{\(\qedsketch\)}
\newcommand{\bpf}{\begin{smallproof}} \newcommand{\epf}{\end{smallproof}}

\newcommand{\BEQ}{\begin{equation}} \newcommand{\EEQ}{\end{equation}}
\newcommand{\BEQN}{\begin{eqnarray}}\newcommand{\EEQN}{\end{eqnarray}}

\renewcommand{\Pr}{{\rm Pr}}

\newcommand{\eps}{\epsilon}

\newcommand{\calD}{{\cal D}}

\renewcommand{\O}{{\rm O}}

\newcommand{\floor}[1]{{\lfloor{#1}\rfloor}}
\newcommand{\todo}[1]{\iffalse {#1} \fi }